\documentclass[12pt]{iopart} 

\input{epsf}
\usepackage{amssymb}
\usepackage{iopams}  

\begin{document}


\title{Hyperbolic orbits of Earth flybys and effects of ungravity-inspired conservative potentials}


\author{O Bertolami$^1$, F Francisco$^1$ and P J S Gil$^2$}
\address{$^1$ Departamento de F\'isica e Astronomia and Centro de F\'isica do Porto, Faculdade de Ci\^encias, Universidade do Porto, Rua do Campo Alegre 687, 4169-007 Porto, Portugal}
\address{$^2$ CCTAE, IDMEC, Instituto Superior T\'ecnico, Universidade de Lisboa, Avenida Rovisco Pais 1, 1049-001 Lisboa, Portugal}

\eads{\mailto{orfeu.bertolami@fc.up.pt}, \mailto{frederico.francisco@icloud.com}, \mailto{p.gil@dem.ist.utl.pt}}

\maketitle


\begin{abstract}	
	In this work we take a critical look at the available data on the flyby anomaly and on the current limitations of attempts to develop an explanation. We aim to verify how conservative corrections to gravity could affect the hyperbolic trajectories of Earth flybys. We use ungravity-inspired potentials as a illustrative examples and show how the resulting orbital simulations differ from the observed anomaly. We also get constraints on the model parameters from the observed flyby velocity shifts. The conclusion is that no kind of conservative potential can be the cause of the flyby anomaly.
\end{abstract}


\maketitle


\section{Introduction}
\label{sec:Intro}

Deep-space probes often use gravity assist manoeuvres to reduce the fuel requirements in order to reach their destinations. In the past few decades, some of those that used Earth gravity assists have shown what appears to be an unexpected shift in their hyperbolic excess velocities. This issue had already been the subject of discussion since the mid 1990's \cite{Antreasian:1998,Anderson:2001}, but had maintained a relatively low profile in the scholarly journals until a paper on the subject by Anderson \textit{et al.}\ came out in 2008 \cite{Anderson:2008}. This was at the height of the Pioneer anomaly controversy that had sparked a few years earlier \cite{Anderson:1998,Anderson:2002}, so the physics community was especially receptive to the discussion of spacecraft trajectory anomalies. The Pioneer anomaly was ultimately solved through conventional heat and radiative momentum transfer mechanisms \cite{Bertolami:2008,Bertolami:2010,Francisco:2012}, with this solution being confirmed by three independent efforts \cite{Francisco:2012,Rievers:2011,Turyshev:2012}. The same can also be stated about the Cassini anomaly observed while that spacecraft was cruising between Jupiter and Saturn, and whose cause is, most likely, also due to radiative momentum transfer \cite{Bertolami:2014}. However, unlike the Pioneer and the Cassini anomalies, this so-called \emph{flyby anomaly} remains an open question ever since.

This anomaly was detected in the residuals of the radiometric tracking data of several space probes performing Earth flybys. The trajectories inferred from the tracking data proved impossible to fit to a single hyperbolic arc. The pre-encounter and post-encounter data had to be fit by separate hyperbolic trajectories that displayed a discrepancy in the hyperbolic excess velocities. Since the two hyperbolic arcs appear to be good fits for their respective data sets, it is assumed that this velocity shift is localised near the perigee, where tracking through the Deep Space Network (DSN) is not available for approximately four hours \cite{Anderson:2008}. This fact has, indeed, led to a proposed experimental setup to improve this time window \cite{Bertolami:2012}. The spatial resolution of the available reconstructions, resulting from the $10~{\rm s}$ interval tracking, does not allow for an accurate characterisation of the effect, so that no corresponding spatial or temporal profile of the acceleration exists. The only available data to support the analysis is the shift in the excess velocity (and correspondingly, in kinetic energy) between the inbound and outbound trajectories.

The flyby anomaly has been observed in the Galileo, NEAR, Rosetta and Cassini flybys. Earth flybys between 1990 and 2005 are listed in Table\ \ref{tbl:flybys}, based on data from Ref.\ \cite{Anderson:2008}, with the respective perigee altitudes ($h_{\rm p}$) and velocities ($v_{\rm p}$), hyperbolic excess velocities ($v_\infty$) and the anomalous shift in the excess velocity $(\Delta v_\infty$).

\begin{table}[htp]
	\begin{center}
		\caption[List of Earth flybys between 1990 and 2005.]{List of Earth flybys between 1990 and 2005, based on data from Ref.\ \cite{Anderson:2008}, where $h_{\rm p}$ and $v_{\rm p}$ denote the altitude and velocity at the perigee, $v_\infty$ is the hyperbolic excess velocity and $\Delta v_\infty$ is the measured change in excess velocity.}
		\begin{tabular}{c c | c c c c}
			\hline
			\hline
			Date		& Mission	& $h_{\rm p}$	& $v_{\rm p}$		& $v_\infty$		& $\Delta v_\infty$		\\
						&			& (${\rm km}$)	& (${\rm km/s}$)	& (${\rm km/s}$)	& (${\rm mm/s}$)		\\
			\hline
			08/12/1990	& Galileo	& $960$			& $13.740$			& $8.949$			& $3.92$				\\
			08/12/1992	& Galileo	& $303$			& $14.080$			& $8.877$			& $-4.6$				\\
			23/01/1998	& NEAR		& $539$			& $12.739$			& $6.851$			& $13.46$				\\
			18/08/1999	& Cassini	& $1175$		& $19.026$			& $16.01$			& $-2$					\\
			04/03/2005	& Rosetta	& $1956$		& $10.517$			& $3.863$			& $1.80$				\\
			02/08/2005	& MESSENGER	& $2347$		& $10.389$			& $4.056$			& $0.02$				\\
			\hline
			\hline
		\end{tabular}
		\label{tbl:flybys}
	\end{center}
\end{table}

It should be noted that the information on the flybys where the anomaly was allegedly detected is not entirely consistent across the different sources. Aside from minor disparities in the values of some parameters, there are a few important differences that deserve some discussion. 

Regarding the second Galileo flyby, Ref.\ \cite{Anderson:2008} explains that the measured value for the change in excess velocity ($\Delta v_\infty$) was actually $-8~{\rm mm/s}$ and the $-4.6~{\rm mm/s}$ figure in Table\ \ref{tbl:flybys} is obtained after subtracting an estimate of the atmospheric drag. However, for the same flyby, two other sources claim that, due to the low altitude, atmospheric drag would mask any anomalous velocity change \cite{Antreasian:1998,Turyshev:2009}. The reasons for absence of any value for the August 1999 Cassini flyby in Ref.~\cite{Turyshev:2009} are also uncertain, as no explanation is offered for this omission.

When one compared the tables presented in Refs.\ \cite{Anderson:2008} and \cite{Turyshev:2009}, one is not merely facing slight differences in values. The two sets of data in these two papers could actually imply a whole different picture on the phenomenon, potentially prompting different assumptions. While in the first the effect appears to be bidirectional, accelerating the spacecraft on some occasions and decelerating in others, in the second reference all instances that have negative velocity changes have been removed and the effect appears to be consistent with an energy increase.

The study of more flybys and closure of the four hour gap in DSN tracking near the perigee would be essential to shed some more light onto this phenomenon. There have been a few recent flybys, listed in Table~\ref{tbl:flybysnodata}, however their results are yet to be presented in scholarly literature.

\begin{table}[htp]
	\begin{center}
		\caption{Recent Earth flybys for which no data is yet available in the literature.}
		\begin{tabular}{c c}
			\hline
			\hline
			Date		& Mission	\\
			\hline
			13/11/2007	& Rosetta	\\
			13/11/2009	& Rosetta	\\
			09/10/2013	& Juno		\\
			03/12/2015	& Hayabusa 2 \\
			\hline
			\hline
		\end{tabular}
		\label{tbl:flybysnodata}
	\end{center}
\end{table}

Some attempts have been made to find explanations for the flyby anomaly, or empirical descriptions like the one put forward in Ref. \cite{Anderson:2002}. A very detailed of the two Galileo (1990 and 1992) and the NEAR (1998) gravity assists considered effects of Earth oblateness, other Solar System bodies, relativistic corrections, atmospheric drag, Earth albedo and infrared emissions, ocean tides and solar pressure \cite{Antreasian:1998}. This analysis was extended to other possible error sources such as the atmosphere, ocean tides, solid tides, spacecraft charging, magnetic moments, solar wind and spin-rotation coupling \cite{Lammerzahl:2008}. None of these efforts was able to find a suitable explanation. Speciffically, the effect of Earth oblateness could yield an acceleration with a compatible order of magnitude, although all attempts yielded unreasonable solutions, unable to account for all flybys \cite{Antreasian:1998}. A study of the thermal effects in the first Rosetta flyby has also been preformed, concluding that they cannot be responsible for the reported flyby anomaly \cite{Rievers:2011}. A discussion about some exotic explanations can be found in Refs. \cite{Bertolami:2012,Francisco:2015th}, including interaction with dark matter \cite{Adler:2009}, modified inertia \cite{McCulloch:2008} or modified particle dynamics \cite{Lammerzahl:2008}. There are other efforts underway, for instance at the IAU, looking into the effects of the definition of coordinate frames on flyby trajectories \cite{Soffel:2003}.

At this point, it must be acknowledged that the flyby anomaly is still poorly characterised and any consistent treatment is, at least, very difficult. These limitations clearly raise the need for alternative approaches. 

In this paper, we first perform an order of magnitude analysis on some mechanical effects that could putatively account for the flyby anomaly. This allows us to acquire some sensitivity on the nature of the phenomenon. We then use an ungravity-inspired potential as an example of a conservative modification to gravity and study its effects on the Earth flyby trajectories. This allows us to reach general conclusions on the effects of conservative potentials on the suspected anomalous hyperbolic trajectories. The main idea to retain is that conservative forces cannot explain the flyby anomaly. 


\section{Newtonian Dynamics}

The data available about the flyby anomaly suggests there may be an energy (and linear momentum) shift highly localised at the perigee. It is sensible to look for conventional means for that shift to take place. This procedure also allows one to gain some sensibility on the figures involved.

We shall thus begin by examining some conventional Newtonian mechanisms, if nothing else, to acquire some sensitivity on the figures of merit. This analysis shall be taken in terms of orders of magnitude.

The most obvious mechanism to induce a change in orbital energy is the separation of a certain amount of mass with a certain speed relative to the spacecraft, in effect, a \emph{thrust}. In this discussion, we shall not be concerned with the specific origin of this mass loss, but instead to attempt to estimate an order of magnitude for the speed and amount of mass that would have to be involved in order to generate a shift in energy similar to the one reported in the flyby anomaly.

The first Galileo flyby of December 8, 1990 provides a good benchmark for an order of magnitude analysis. It had a perigee altitude of approximately $10^3~{\rm km}$ with a maximum velocity of around $14~{\rm km/s}$. Its hyperbolic excess velocity was $v_{\infty} \approx 9~{\rm km/s}$, to which corresponds a specific orbital energy (energy per unit mass) $\mathcal{E} \approx 4 \times 10^7~{\rm J/kg}$.

The frequency shift implied by the residuals for this flyby could be explained by an increase in hyperbolic excess velocity $\Delta v_{\infty} \approx 4~{\rm mm/s}$ or, equivalently, a shift in specific energy $\Delta \mathcal{E} \approx 35~{\rm J/kg}$. This corresponds to a $\Delta v \approx 2.6~{\rm mm/s}$ at the perigee.

Assuming that this $\Delta v$ is the result of some kind of thrust, we can, from basic mechanics, through conservation of linear momentum, obtain a relation between the mass and speed of the propellant. It is not relevant for this analysis to speculate if this propellant is a solid fragment or a gas. We consider the most favorable case of thrust along the flight direction yielding the largest possible effect an thus obtaining an upper bound of this effect. Also, for simplicity, we assume the thrust to be instantaneous, which seems reasonable given the highly localised nature of the anomaly we are searching for.

The results we find from this analysis show that any fragment or amount of mass small enough to be accidentally lost without being detected, would have to have a speed relative to the spacecraft itself of, at least, the same order of magnitude as the spacecraft's velocity relative to Earth. Indeed, for a fragment or leak small enough for an undetected loss, let us say of around $10^{-7}$ of the mass of the spacecraft or $0.25~{\rm g}$, the required relative speed would be of the order of $10^4~{\rm m/s}$ for an ejection along the direction of the motion, the most favourable angle. 

Even though there is no known internal mechanism for such thing to happen without being detected, one could consider the reverse process. If the spacecraft were to collide with a piece of debris or a meteoroid, relative speeds of this order are, indeed, possible. If the collisions were to take place at the right angles, this could be a possibility to obtain the kinds of velocity shifts obtained, although the spacecraft could sustain signifficant damage. However, a recently published survey of damage sustained by the Space Shuttle from micrometeoroids and orbital debris has found that the most frequent damaging fragments have masses ranging from $10^{-5}~{\rm g}$ to $10^{-3}~{\rm g}$ \cite{Hyde:2015}, which is at least two orders of magnitude below what we used in our estimate, for comparable speeds. Meteoroids with the required mass must be rare enough to be ruled out as an explanation for the flyby anomaly.

Another speculation to achieve the reported energy shift is a change in rotational kinetic energy resulting from a transfer of translational kinetic energy, with the consequent coupling between linear and angular momentum. Again, we do not attempt to propose a specific mechanism for this, but only to assess the viability of such a putative process through an estimate of the involved orders of magnitude of the energy exchange.

From the approximate dimensions of these typical spacecraft (\textit{e.g.} Galileo is a $\sim 300~{\rm kg}$ spacecraft with an approximately cylindrical shape with a $\sim 1~{\rm m}$ radius, a $\sim 3~{\rm m}$ height), one can estimate the moment of inertia of the spacecraft and compute the change in angular velocity due to the translational energy change. A typical figure of merit is around $10~{\rm rad/s}$, which is clearly too large to go undetected. Thus, one can conclude that an explanation for the flyby anomaly from any kind of rotational mechanism is unfeasible. 

This analysis, even though straightforward and approximate, alows for acquiring some sensitivity on the figures involved in the physical processes, from a classical mechanical standpoint. Hence, we verify that, despite being a small effect from the orbital analysis point of view, the flyby anomaly involves a shift in velocity and energy that is too large to be caused by any kind of subtle hypothetical mechanism.


\section{An example of a conservative modification to gravity: Ungravity-inspired potential}
\label{sec:UngravPot}


\subsection{Unparticles}

Recently, the possibility of the existence of new physics above the TeV scale has been considered through the introduction of unparticles \cite{Georgi:2007,Georgi:2007a}. In this scheme one admits a hidden sector with a nontrivial infrared fixed point $\Lambda_{\rm U}$, below which scale invariance is explicit. In the ultraviolet (UV) regime, at energies above $\Lambda_{\rm U}$, the hidden sector operator $\mathcal{O}_{\rm UV}$ of dimension $d_{\rm UV}$ couples to the standard model (SM) fields through an operator $\mathcal{O}_{\rm SM}$ of dimension $n$ via nonrenormalizable interactions $\mathcal{O}_{\rm UV} \mathcal{O}_{\rm SM} / M_{\rm U}^{d_{\rm UV}+n-4}$, where $ M_{\rm U}$ is the mass of the heavy exchanged particle. Below $\Lambda_{\rm U}$, the hidden sector becomes scale invariant and the operator $\mathcal{O}_{\rm UV}$ mutates into an unparticle operator $\mathcal{O}_{\rm U}$ with noninteger scaling dimension $d_{\rm u}$. The coupling of field operators can be generically written as
\begin{equation}
	{\Lambda_{\rm U}^{d_{\rm UV}-d_{\rm u}} \over M_{\rm U}^{d_{\rm UV}+n-4}} \mathcal{O}_{\rm U}\mathcal{O}_{\rm SM}
\end{equation}
The operator $\mathcal{O}_{\rm U}$ could be a scalar, a vector, a tensor or even a spinor.

The exchange of unparticles gives rise to long range forces which deviate from the inverse-square law (ISL) for massless particles due to the anomalous scaling of the unparticle propagator. For example, the exchange of scalar (pseudo-scalar) unparticles can give rise to spin-dependent long range forces, as pointed out in Ref. \cite{Liao:2007}. Coupling between unparticles and vector or axial-vector currents have been investigated in Ref. \cite{Deshpande:2008}. In Ref. \cite{Goldberg:2008} the coupling between unparticles and the energy-momentum tensor was studied. An analysis of some of the phenomenological implications can be found in Refs.\ \cite{Bertolami:2009,Bertolami:2009a,Alves:2010}.


\subsection{Ungravity}

If $\mathcal{O}_{\rm U}$ is a rank-2 tensor, it can couple with the stress-energy tensor $T_{\mu\nu}$ and lead to a modification to Newtonian gravity. Taking the gravitational coupling of the tensor unparticle to $T_{\mu\nu}$ to be of the form
\begin{equation}
	{1 \over M_{\star} \Lambda_{\rm U}^{d_{\rm u}-1}}\sqrt{g}T_{\mu\nu}\mathcal{O}_{\rm U}^{\mu\nu}
\end{equation}
where $M_{\star} = \Lambda_{\rm U}(M_{U}/\Lambda_{\rm U})^{d_{\rm UV}}$, it can be shown, in the non-relativistic limit and for $d_{\rm u} \neq 1$, that the effective gravitational potential with the unparticle exchange has the form \cite{Goldberg:2008}
\begin{equation}
	V(r) = -{G_{\rm N} M \over r} \left[1 + \left({R_{\rm G} \over  r}\right)^{2d_{\rm u}-2} \right],
	\label{eq:UngravPotTens}
\end{equation}
where $G_{\rm N}$ is the Newtonian gravitational constant and $R_G$ is a characteristic length scale.

One can also obtain a force from the coupling of a vector unparticle \cite{Deshpande:2008}. The potential for the coupling between a vector unparticle and a baryonic (or leptonic) current $J_{\mu}$ of the form
\begin{equation}
	{\lambda \over \Lambda_{\rm U}^{d_{\rm u}-1}}J_{\mu}\mathcal{O}_{\rm U}^{\mu}
\end{equation}
is given, when combined with the gravitational potential, by
\begin{equation}
	V(r) = -{G_{\rm N} M \over r} \left[1 - \left({R_{\rm G} \over  r}\right)^{2d_{\rm u}-2} \right].
	\label{eq:UngravPotVec}
\end{equation}

In this paper, we use Eqs.~(\ref{eq:UngravPotTens}) and (\ref{eq:UngravPotVec}) from a phenomenological model standpoint, and we shall refer to them, respectively, as ``tensorial'' and ``vectorial'' ungravity-inspired potentials. In this context, we attempt to constrain the values of $R_{\rm G}$ and $d_{\rm u}$ from the velocity and energy shifts observed in the Earth flybys discussed in Section \ref{sec:Intro}. This is analysed from two different perspectives. First, we look at the way the hyperbolic trajectories of these spacecraft would be altered by the existence of an ungravity-inspired potential and, particularly, what is the temporal signature of the radial velocity perturbation. In a second analysis, we attempt to obtain the range of  values for the potential paramenters $R_{\rm G}$ and $d_{\rm u}$ that leads to a velocity anomaly that fits the order of magnitude of the observed instances.


\section{Results and Discussion}

The existence of a potential of the kind shown in Section \ref{sec:UngravPot} would have observable effects on the trajectories of objects under gravity. Specifically, the hyperbolic trajectories of the Earth flybys under analysis in this paper would show a deviation from their predicted trajectories.

In this Section, we aim to characterise the perturbation induced by an ungravity-inspired potential of the form shown in Eqs.~(\ref{eq:UngravPotTens}) and (\ref{eq:UngravPotVec}). In order to achieve that, we have performed numerical simulations of the orbits reproducing the conditions of the flybys and repeated them with the addition of the new potential. We can then compare the two trajectories in the search for the observational signature of the perturbation and the bounds on the parameters established by the flyby data.


\subsection{Temporal Signature}

The anomalous velocity shift that was discovered in some Earth flybys is inferred from the inability to fit the inbound and outbound arms of the trajectory to a single hyperbolic segment. It shows up in the Doppler data residuals as a sudden step in frequency at the perigee.

The Doppler frequency shift is a function of the radial velocity of the probe. For this reason, the easiest way to look for a Doppler frequency shift from a trajectory simulation is to extract the radial velocity perturbation resulting from the new potential and look at its time signature. This is a similar methodology to the one used in Ref.~\cite{Iorio:2014}.

At this stage, we will not look at the values of the parameters, though some preliminary simulations have shown us that for values of $R_{\rm G}$ near the Earth radius, $R_{\oplus}$, the order of magnitude of $d_{\rm u}$ must be within $1 \pm 10^{-6}$ in order to produce a velocity shift that is remotely compatible with the ones reported in the flyby anomaly.

\begin{figure}
	\centering
	\epsfxsize= 0.7\columnwidth
	\epsffile{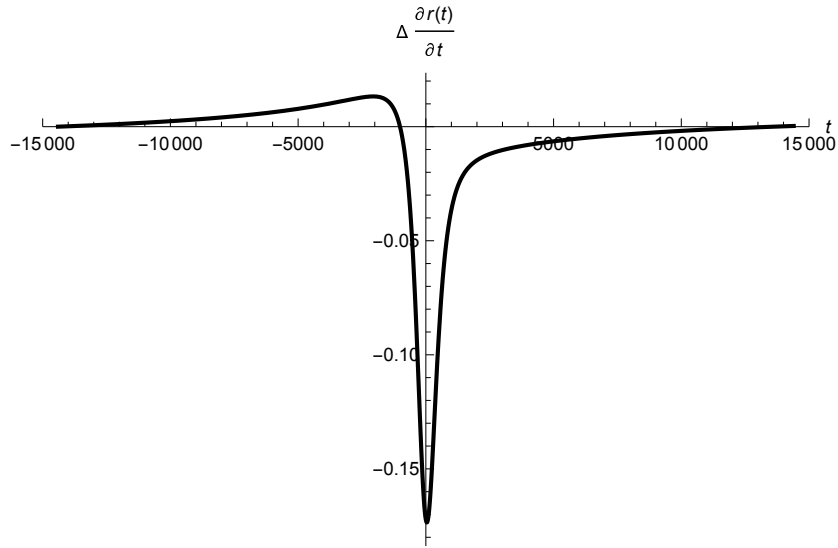}
	\caption{Difference between the radial velocity of the hypothetical trajectory with a tensorial ungravity-inspired potential and the Newtonian trajectory as a function of time. Both trajectories are simulated for the same initial conditions. The tensorial ungravity-inpired potential uses $R_{\rm G} = R_{\oplus}$ and $d_{\rm u} = 1 + (5 \times 10^{-7})$,}
	\label{TimeSignT}
\end{figure}

\begin{figure}
	\centering
	\epsfxsize= 0.7\columnwidth
	\epsffile{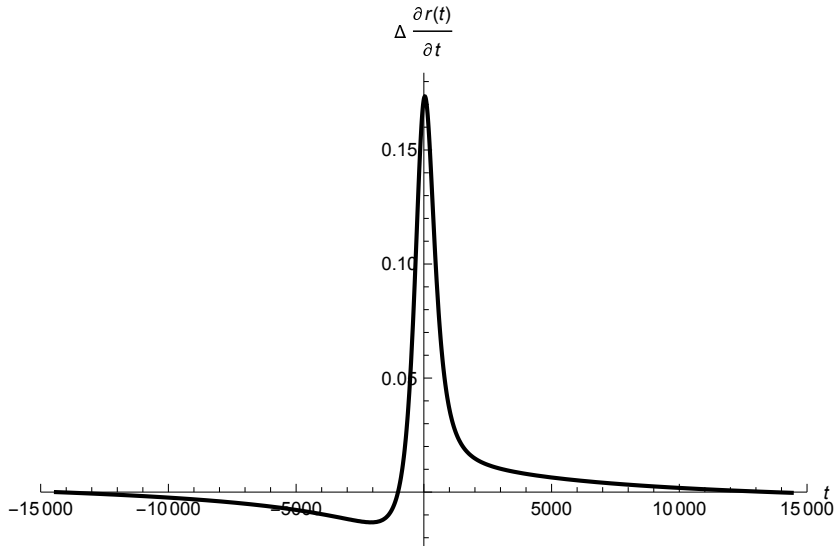}
	\caption{Same as Fig.\ \ref{TimeSignT}, for a vectorial ungravity-inspired potential, with $R_{\rm G} = R_{\oplus}$ and $d_{\rm u} = 1 + (5 \times 10^{-7})$.}
	\label{TimeSignV}
\end{figure}

When we plot the perturbation in radial velocity relative to the Newtonian trajectory as a function of time, the most striking feature is a large spike localised between approximately 10 minutes before and after the perigee, as can be seen in Figs.\ \ref{TimeSignT} and \ref{TimeSignV} . Besides that, there is a smaller deviation in the inbound and outbound trajectories, but that converges to the Newtonian value as we get further from the perigee. Figs.\ \ref{TimeSignT} and \ref{TimeSignV} also show the difference between applying a ``tensorial'' or ``vectorial'' unparticle potential, which basically flips the graphic in the vertical direction. 

From the observation of these results, it is already obvious that an ungravity-inspired potentials cannot produce a perturbation that would yield the kind of temporal signature that is observed in the Doppler residuals of the Earth flybys. Still, one might be tempted to reason that, if the large spike around the perigee would somehow not appear in observational data due to lack of resolution, what remains is a step in radial velocity that somewhat resembles what is observed in the residuals \cite{Anderson:2008}. One should, thus, look closely at Figs.\ \ref{TimeSignT} and \ref{TimeSignV} and observe, once again, that the velocity shift quickly converges to zero on both the descending and ascending trajectories as one gets further away from the perigee, a behaviour consistent with energy conservation, unlike the flat behaviour that is observed in the residuals that implies a non-conservative effect. Furthermore, the way in which these time signatures were obtained means that they are not directly comparable to the Doppler residuals.

We can stress this point further by attempting to reproduce the conditions in which those residuals were produced. To do that, we now compare two Newtonian hyperbolic arms obtained separately from the inbound and outbound parts of an underlying hypothetical trajectory with the ungravity-inspired potential. 

In this case the results show a similar spike in the radial velocity shift near the perigee, although the overall time signature is now symmetric around the perigee. This last feature is due to the fact that the perturbation to the potential is still an inverse power law of the radius. This means that the integration of the perturbation along any time interval centred at the perigee will always lead to a symmetric time signature converging to zero at the extremities.

\begin{figure}
	\centering
	\epsfxsize= 0.7\columnwidth
	\epsffile{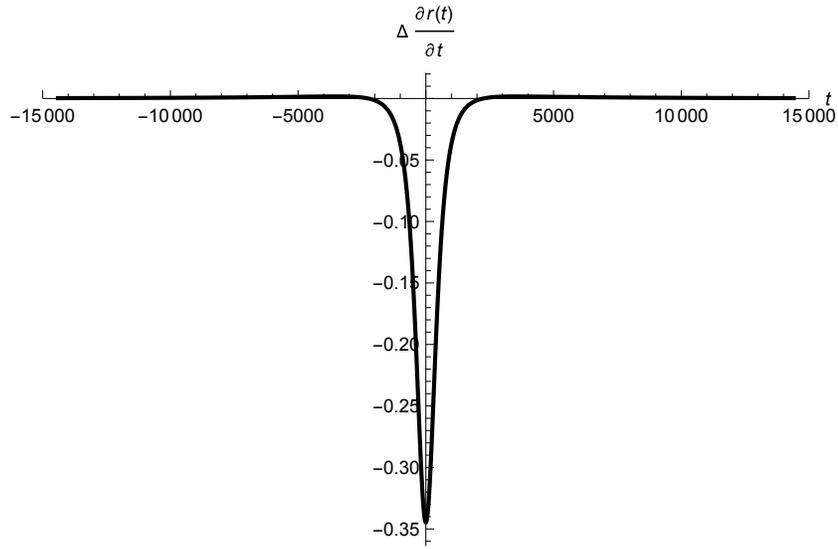}
	\caption{Difference between the radial velocity of the outbound and inbound Newtonian trajectory arms computed from the underlying hypothetical trajectory perturbed with a tensorial ungravity-inspired potential as a function of time. This process aims to reproduce the way in which residuals are obtained in Ref.\ \cite{Anderson:2008}. The tensorial ungravity-inpired potential uses $R_{\rm G} = R_{\oplus}$ and $d_{\rm u} = 1 + (5 \times 10^{-7})$, and the Newtonian trajectory.}
	\label{ResidualsT}
\end{figure}

\begin{figure}
	\centering
	\epsfxsize= 0.7\columnwidth
	\epsffile{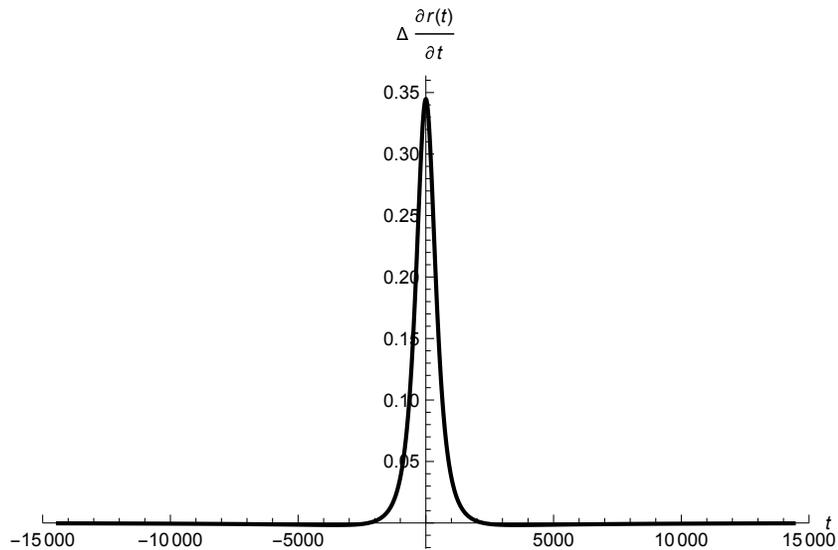}
	\caption{Same as Fig.\ \ref{ResidualsT}, for a vectorial ungravity-inspired potential, with $R_{\rm G} = R_{\oplus}$ and $d_{\rm u} = 1 + (5 \times 10^{-7})$.}
	\label{ResidualsV}
\end{figure}

From this last observation, it is logical to conclude that, since any kind of potential based on position dependent deviation to the inverse square law will be unable to explain the flyby anomaly. One can reinforce this argument by recalling that a potential that preserves energy conservation can only depend on the coordinates. Since the velocity shift in the flyby anomaly implies a corresponding orbital energy shift, it cannot be explained by any kind of conservative force.


\subsection{Constraints on $R_{\rm G}$ and $d_{\rm u}$}

Although it is now clear that the temporal signature of ungravity-inspired potentials are not compatible with the one reported in the flyby anomaly, it is still interesting to attempt to use the velocity shifts involved in these flybys to constrain the ungravity parameters $R_{\rm G}$ and $d_{\rm u}$.

To bound these parameters, we had to perform a series of numerical simulations where the maximum radial velocity shift caused by the ungravity potential is measured. We then plot the maximum velocity shift as a function of $R_{\rm G}$ and $d_{\rm u}$. Due to the characteristic perigee altitudes involved in these flybys, we look for characteristic length scales near the Earth radius, \textit{i.e.}, $R_{\rm G} \sim R_{\oplus}$. A preliminary set of simulations also indicated that the values of $d_{\rm u}$ should be within $10^{-6}$ of unity.

\begin{figure}
	\centering
	\epsfxsize= 0.8\columnwidth
	\epsffile{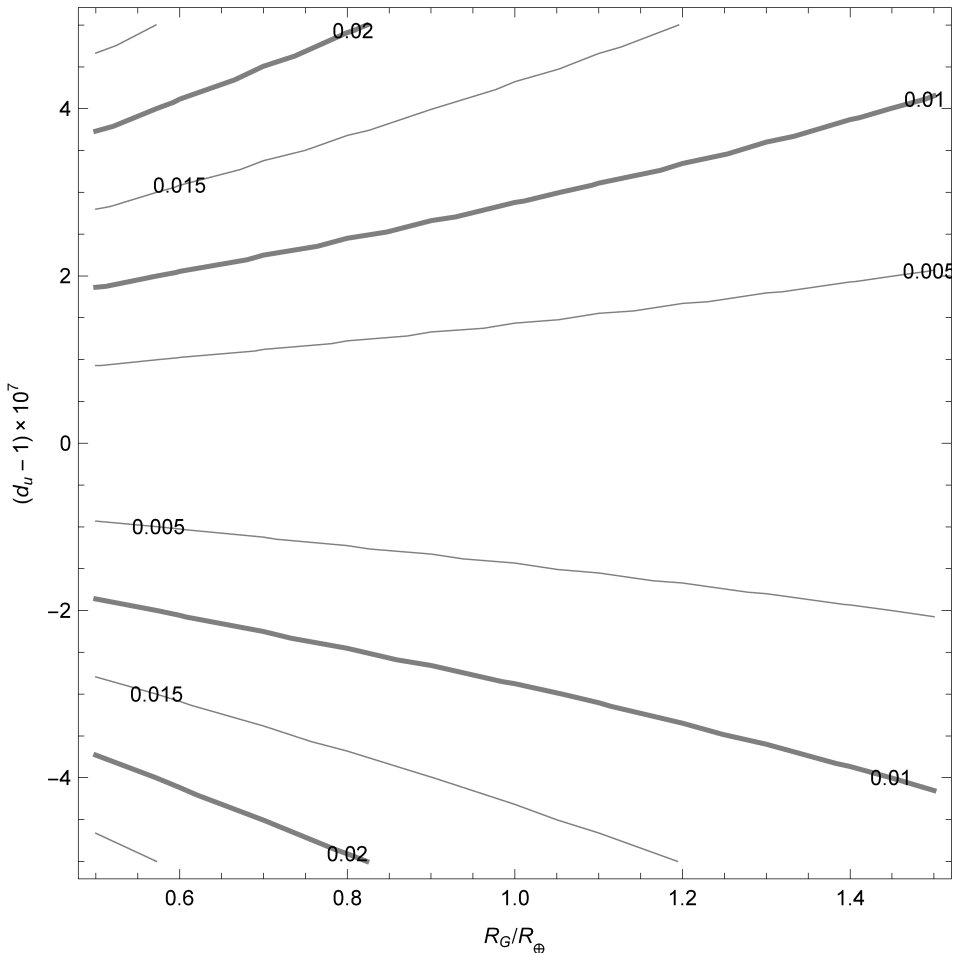}
	\caption{Contour plot of the maximum radial velocity shift, in $\rm m/s$, of the trajectory perturbed by the ungravity potential relative to the Newtonian trajectory, as a function of $R_{\rm G}$ and $d_{\rm u}$. The plot is similar for tensorial and vectorial potentials.}
	\label{Parameters}
\end{figure}

The result is depicted in Fig.\ \ref{Parameters} where the dependence of the maximum perturbation shift on the ungravity parameters is shown. In this case, we tested values of $d_{\rm u}$ below and above unity, though always very close to it, as discussed above. The results are similar for tensorial and vectorial ungravity, since, as we have also discussed, the difference between these two only causes the vertical flipping of the time signatures, which is the same effect that the change from $d_{\rm u} \gtrsim 1$ to $d_{\rm u} \lesssim 1$ has.

What these results show is that the perturbation grows as $d_{\rm u}$ is further away from $1$, as would be expected. The same can be stated as $R_{\rm G}$ gets smaller. More importantly, we can see that the range of values that is compatible with a velocity shift of the order of $10~{\rm mm/s}$ or less is a narrow strip around $d_{\rm u} = 1$, that gets wider as $R_{\rm G}$ grows. Still, a much larger $R_{\rm G}$ would start to affect other orbits at larger scales and would not yield a localised effect like the flyby anomaly.

Overall, we stress once again that, not only do the flyby anomaly measurements place very stringent limits on $d_{\rm u}$, but also that this parametric analysis does not take into account the above discussion about the time signature of the effect, where we pointed out that no conservative effect can alone explain the flyby anomaly.

As a final remark, we could point out that, if the effect of a conservative force to be added to the Newtonian component changes the perigee altitude, then the importance of dissipative forces such as atmospheric drag might be expected to change as well, as density has an exponential dependence with altitude. However, for the typical altitudes involved, the density is small enough so that the effect of atmospheric drag on the velocity is already much smaller that the anomaly. Clearly, this is not relevant for the cases depicted in Table \ref{tbl:flybys}, but the fact remains that the only possibility for a conservative force to be relevant in the flyby anomaly is through a coupling with some kind of non-conservative force.


\section{Conclusions}

The flyby anomaly has been a puzzle for quite some time and, so far, no credible explanation has emerged. This paper does not provide a final answer either, but eliminates some possible causes and raises attention for the time signature test that any possible explanation of the flyby anomaly must pass.

Given the discussion we have made in Section\ \ref{sec:Intro}, a significant effort must still be made in what concerns the characterisation of the phenomenon. It would be extremely important to have results on the trajectory analysis for the most recent flybys. Otherwise, the six observational instances on which this whole discussion is based, with no known independent analyses and with all the caveats that we have discussed, may not be sufficient to unequivocally establish the features that are attributed to this anomaly, or indeed its existence. The proposal of Ref.\ \cite{Bertolami:2012} can provide a relevant contribution in this direction.

The discussion around these ungravity-inspired potentials sets out another example of an attempt to tackle this anomaly based on a modified gravity model. It gives a characterisation of the time signature that this kind of modification to Newtonian gravity yields. For the parameter range tested in this paper, namely, for length scales similar to the Earth radius, we obtain a highly localised effect near the perigee. Still, the effect is symmetric around the perigee, with the perturbation converging back to zero as the object goes further from the Earth flyby, as would be expected for any conservative force. We also constrained the length scale and non-integer dimension of the ungravity potential with bounds set by the flybys where the anomaly was detected, leading to a very narrow range of allowed values for this last parameter.

In the end, we must conclude that an potential inspired by unparticle-like corrections, such as the ones used in this work, cannot explain the flyby anomaly. Indeed, the observations made about its time signature remain valid for any inverse power law of distance which necessarily conserve orbital energy. The flyby anomaly has an inherently non-conservative nature.

The search for a solution for the flyby anomaly has been akin to the work of a detective, and has yet to provide any palpable results or, indeed, significant advances in the knowledge of the anomaly itself. Until more observational data is available, we predict that the situation is unlikely to change.


\section*{Acknowledgments}

The work of F.\ Francisco has been partially sponsored through a fellowship from Faculdade de Engenharia da Universidade do Porto, under the supervision of Prof. Jaime Villate.

The work of P.J.S.\ Gil was supported by FCT, through IDMEC, under LAETA, project UID/EMS/50022/2013.


\section*{References}

\bibliography{flyby_ungrav}

\end{document}